\documentclass[
    ,final            
  ]
  {aipproc}

\def\sles{\lower2pt\hbox{$\buildrel {\scriptstyle <} 
   \over {\scriptstyle\sim}$}}
\def\sgreat{\lower2pt\hbox{$\buildrel {\scriptstyle >} 
   \over {\scriptstyle\sim}$}}

\layoutstyle{6x9}

\begin{document}

\title{Evidence for the Black Hole Event Horizon}

\author{Ramesh Narayan$^*$}{
  address={Institute for Advanced Study, Princeton, NJ 08540},
  address={Harvard-Smithsonian Center for Astrophysics, Cambridge, MA 02138}
}

\author{Jeremy S. Heyl}{
  address={Harvard-Smithsonian Center for Astrophysics, Cambridge, MA 02138}
}

\begin{abstract}
Roughly a dozen X-ray binaries are presently known in which the
compact accreting primary stars are too massive to be neutron stars.
These primaries are identified as black holes, though there is as yet
no definite proof that any of the candidate black holes actually
possesses an event horizon.  We discuss how Type I X-ray bursts may be
used to verify the presence of the event horizon in these objects.
Type I bursts are caused by thermonuclear explosions when gas accretes
onto a compact star.  The bursts are commonly seen in many neutron
star X-ray binaries, but they have never been seen in any black hole
X-ray binary.  Our model calculations indicate that black hole
candidates ought to burst frequently if they have surfaces.  Based on
this, we argue that the lack of bursts constitutes strong evidence for
the presence of event horizons in these objects.
\end{abstract}

\maketitle


\section{Introduction}

X-ray binaries \cite{1995xrb..book.....L} are double star systems in
which mass is transferred from a secondary star onto a compact primary
star.  The gas radiates X-rays as it accretes onto the primary, which
is either a neutron star (NS) or a black hole (BH).  Of the over 200
X-ray binaries known in the Galaxy, about 30 are transient sources,
known as ``X-ray novae'' or ``soft X-ray transients''
\cite{TS96,vPM95}.  These sources are characterized by episodic
outbursts at X-ray, optical and radio frequencies, separated by long
intervals (years to decades) of quiescence.  X-ray novae have played a
major role in the hunt for BHs.

In the quiescent state of an X-ray nova, the absorption-line
velocities of the secondary star can be measured precisely because the
non-stellar light from the accretion flow is modest compared to the
light of the secondary.  From a measurement of the velocity of the
secondary as a function of the binary orbital phase, one may directly
determine the mass function,
\begin{equation}
f(M) \equiv {M_{1}^{3}\sin^{3}i \over (M_{1}+M_{2})^{2}},
\end{equation}
where $M_{1}$ and $M_{2}$ are the masses of the primary and secondary
star, and $i$ is the orbital inclination angle.  The mass function
provides a strict lower limit on the mass of the primary.

In nearly a dozen X-ray novae with reliable measurements of the mass
function, it has been found that $f(M)\ \sgreat\ 3M_{\odot}$ (see
\cite{NGM01} for a list of X-ray novae with large mass functions as of
mid-2001, and \cite{GCM01} for a newly confirmed system).  Assuming
that general relativity applies, the maximum mass of a NS can be
calculated to be no more than about $2-3M_\odot$
\nocite{RR74,CST94,KB96}[6--8].
Therefore, there is strong reason to think
that the compact primaries in these X-ray novae are BHs.

While the mass criterion is an excellent technique for finding BH
candidates, one would like to have more direct evidence that the
candidates truly are BHs.  Ideally, one would like to be able to show
that a candidate BH has an {\it event horizon}.  This is the topic of
the present article.

The outburst in an X-ray nova is caused by a sudden increase in the
mass accretion rate onto the compact primary.  The X-ray flux rises on
a time scale of days, and subsequently declines on a time scale of
weeks or months.  Over the course of the outburst, the mass accretion
rate varies by orders of magnitude, which makes these systems
particularly useful for studying the nature of the central star.
Basically, one can monitor the response of the star to a wide range of
accretion rates, and thereby obtain a more complete understanding of
the properties of the star than one could with a steadily accreting
star.

Roughly half the known X-ray novae contain BH candidates and the rest
contain NSs.  X-ray novae with NSs have X-ray luminosities in
quiescence of the order of $10^{-5} -10^{-6}L_{Edd}$, where $L_{Edd}$
is the Eddington luminosity.  X-ray novae with BH candidates are much
dimmer, with quiescent luminosities in some cases below
$10^{-8}L_{Edd}$ \cite{Getal01}.  The anomalous dimness of quiescent
BH candidates has been used to argue for the presence of event
horizons in these objects [\nocite{Getal01,NGM97,Metal99} 9--11, see
\nocite{NGM01} 4 for a review].  The argument is in essence that black
holes are significantly {\it blacker} than neutron stars and must
therefore have event horizons.

We discuss in this paper a different method of testing for the
presence of event horizons in BH X-ray binaries.  This method again
makes use of the fact that the accreting primary in an X-ray nova
experiences a wide range of mass accretion rates.  However, instead of
focusing on the luminosity of the object in quiescence, we study the
presence or absence of Type I X-ray bursts.

When gas accretes onto a compact star such as a NS, it is compressed
and heated as it accumulates on the surface, leading to thermonuclear
reactions.  In many NS systems (both X-ray novae and other types of NS
X-ray binaries), the reactions occur unsteadily and cause sudden brief
bursts in the X-ray flux \cite{G76}.  The flux rises almost to the
Eddington level within about a second and decays exponentially over a
few seconds or tens of seconds.  These thermonuclear bursts are called
Type I bursts (to distinguish them from Type II bursts, which are not
thermonuclear and are observationally distinct \cite{HML78}).  Type I
bursts have been observed in a large number of NS systems 
\citep[see][for a review]{LvPT93}, and the theory of the bursts is
relatively well understood \nocite{LvPT93,HvH75,WT76,J77,TP78,FHM81,
P82,FHIR84,FHIR87,FL87b,TWL96,B98}[14--25].

No Type I burst has been seen in any BH X-ray binary, even though, as
we show below, BH candidates ought to produce bursts as efficiently as
NSs if they possess surfaces.  We argue that the lack of bursts
represents strong evidence for the presence of event horizons.

\section{A Simple Model of Type I Bursts}

We have developed a simple model to investigate the stability of gas
accreting on the surface of a compact star.  (The discussion in this
section and the next closely follows ref. \cite{NH02}.)  We consider a
compact spherical star of mass $M$ and radius $R$, accreting gas
steadily at a rate $\dot\Sigma ~({\rm g\,cm^{-2}\,s^{-1}})$.  In the
local frame, the gravitational acceleration is $g=GM(1+z)/R^2$, where
the redshift $z$ is given by $1+z=(1-R_S/R)^{-1/2}$, and $R_S=2GM/c^2$
is the Schwarzschild radius.  We assume that the accreting material
has mass fractions $X_0$, $Y_0$ and $Z_0=1-X_0-Y_0$, of hydrogen,
helium and heavier elements (mostly CNO).  The underlying star, as
well as the fully burnt material sitting on it, is taken to have a
composition $X=Y=0$, $Z=1$.

We consider a layer of accreted material of surface density
$\Sigma_{max}$ sitting on top of a substrate of fully burnt material.
Since the physical thickness of the layer is much less than the
radius, we work in plane parallel geometry and take $g$ to be
independent of depth.  We solve for the density $\rho$, the
temperature $T$, the outgoing flux $F$, and the hydrogen, helium and
heavy element fractions, $X$, $Y$, $Z=1-X-Y$, as functions of the
column density $\Sigma$.

Hydrostatic equilibrium, combined with the equation of state
$P=P(\rho,T)$ of the gas, gives
\begin{equation}  
{\partial P\over \partial\Sigma}={\partial
P\over\partial\rho}{\partial\rho\over \partial\Sigma} +{\partial
P\over\partial T}{\partial T\over
\partial\Sigma}=g. \label{hydrostatic}
\end{equation}
For the equation of state we use expressions given in \cite{P82} for
the gas, radiation and degeneracy pressure, modified as needed when
the degenerate electrons are relativistic.

H- and He-burning are described by two differential equations:
\begin{equation}
{dX\over dt}=-{\epsilon_H\over E_H^*}, \quad {dY\over dt}=-{dX\over
dt} -{\epsilon_{He}\over E_{He}^*}, \qquad {d \over dt} \equiv
{\partial \over \partial t} + \dot\Sigma {\partial \over
\partial\Sigma} .
\label{XY}
\end{equation}
Here, $\epsilon_{H}$, $\epsilon_{He}$ are the nuclear energy
generation rates for H and He burning, and $E_{H}^*$, $E_{He}^*$ are
the corresponding energy release per unit mass of H and He burned
\cite{P82}.  For $\epsilon_H$, we include the pp chain and the CNO
cycle, including fast-CNO burning, saturated CNO burning, and electron
capture reactions, as described in \cite{MD84,BC98}.  Since we are not
concerned with modelling the bursts themselves, and since our
stability criterion does not depend on the detailed treatment of the
deep crust, we do not include proton captures onto heavier nuclei (see
\cite{Setal99} for a discussion of some of the consequences of the
$rp$-process burning on accreting NSs).  For He-burning, we include
the triple-$\alpha$ reaction, but not pycnonuclear reactions
\cite{ST83}; the latter are important only at greater pressures than
we consider.  We do not correct the reaction rates to include
screening since we are concerned only with determining whether nuclear
burning of H and He can proceed stably under given conditions; stable
burning of H and He utilizes almost exclusively the reactions included
in our model.

Radiative transfer gives another differential equation:
\begin{equation}
{\partial T \over \partial\Sigma}={3\kappa F \over 4acT^3}, \qquad
{1\over\kappa} = {1\over\kappa_{rad}}+{1\over\kappa_{cond}} .
\label{radtran}
\end{equation}
We employ the fitting functions in \cite{I75} for the radiative
opacity $\kappa_{rad}$, and an analytical formula from \cite{C68},
suitably modified for relativistic electrons, for the conductive
opacity $\kappa_{cond}$; the formula for the latter agrees well with
more modern treatments \cite[e.g.][]{HH01}.

Finally, the energy equation gives
\begin{equation} 
\rho T{ds\over dt} = \rho(\epsilon_H+\epsilon_{He}) + \rho{\partial
F\over \partial\Sigma}, \label{energy}
\end{equation} 
where $s$ is the entropy per unit mass.  The above five differential
equations (2)--(5) form a closed set for the five variables, $\rho$,
$T$, $F$, $X$, $Y$.

We need five boundary conditions to solve the equations.  We apply
four boundary conditions at the surface of the star ($\Sigma=0$) and
one at the base of the accretion layer ($\Sigma=\Sigma_{max}$).  Two
of the four surface conditions are obvious: $X=X_0, ~Y=Y_0$.  We
obtain the third boundary condition by equating the accretion
luminosity of the infalling gas, $L_{acc}=4\pi R^2\dot\Sigma
c^2z/(1+z)$, to blackbody emission from the surface: $4\pi R^2\sigma
T_{out}^4=L_{acc}$.  This fixes the surface temperature $T_{out}$.
Then, using $T_{out}$ and an assumed value of $F_{out}$, we solve for
the surface density profile $\rho(\Sigma)$ from the radiative transfer
equation, thus obtaining the fourth boundary condition.

At the base of the accretion layer we have an inner boundary
condition, which plays an important role in the problem.  For the
calculations presented here, we assume that the base temperature,
$T_{in}=T(\Sigma_{max})$, is fixed.  We examine several values of
$T_{in}$ for layers with $\Sigma_{max}=10^9, 10^{10}$ and $10^{11}$~g
cm$^{-2}$.  Applying the boundary condition at $\Sigma_{max}$ rather
than deeper in the substrate is an approximation, but the error due to
this is not serious.  For the high surface densities we consider, the
heat transfer is dominated by conduction \cite{HH01}, so the
temperature gradient for $\Sigma>\Sigma_{max}$ is small.  Moreover, we
examine models for several values of $T_{in}$ which further mitigates
any error.

The calculations proceed in two stages.  First, we solve for the
steady state profile of the accretion layer by setting
$\partial/\partial t=0$ and replacing the operator
$\partial/\partial\Sigma$ in equations (2)--(5) by $d/d\Sigma$.  With
this substitution, we have five ordinary differential equations with
four outer boundary conditions and one inner boundary condition.  The
solution to these equations gives the profiles of the basic fluid
quantities: $\rho(\Sigma), ~T(\Sigma), ~F(\Sigma), ~X(\Sigma),
~Y(\Sigma)$.

Having calculated the steady state structure of the accretion layer,
we next check its stability.  Various local stability criteria have
been discussed in the literature \cite{B98}, in which one considers
the properties of the gas at a single depth.  While this approach is
very useful for physical insight, it is clearly inadequate for
quantitative work.  Occasionally, some authors have considered a
global criterion involving an integral over the entire layer
\cite{FL87b}.  Even this approach is not very satisfactory since it
usually involves a restriction on the perturbations (e.g., constant
temperature perturbation in \cite{FL87b}).

We have carried out a full linear stability analysis of the accretion
layer.  To our knowledge, this is the first time that this has been
attempted in the field of Type I bursts.  We start with the steady
state solution described above and assume that it is slightly
perturbed, $Q(\Sigma) \to Q(\Sigma) +Q'(\Sigma)\exp(\gamma t)$, where
$Q$ corresponds to each of our five variables, and the perturbations
$Q'(\Sigma)$ are taken to be small.  We linearize the five partial
differential equations given in equations (2)--(5) (three of which
involve time derivatives), apply the boundary conditions, and solve
for the eigenvalue $\gamma$.  We obtain a large number of solutions
for $\gamma$, many of which have both a real and an imaginary part.
We consider the accretion layer to be unstable if any eigenvalue has a
real part (growth rate) greater than the characteristic accretion rate
$\gamma_{acc}=\dot\Sigma/\Sigma_{max}$.

If the steady-state model is unstable according to the linear
stability analysis, accretion cannot proceed stably with the
particular $\Sigma_{max}$ and $\dot\Sigma$.  Whether this instability
manifests itself as a Type~I burst depends on how the burning flame,
once ignited at some random point, envelopes the surface of the star.
This is as yet an unsolved problem, although ref. \cite{SLU02}
discusses a possible solution.  In the following, we assume that the
instability will grow rapidly to the nonlinear regime and that the
system will exhibit Type I bursts.

\section{Results}

\subsection{Type I Bursts on Neutron Stars}

\begin{figure}
\caption{Regions of instability, shown by dots, as a function of
accretion luminosity and stellar radius.  Top left: $1.4M_\odot$ NS
with a base temperature $T_{in}=10^{8.5}$~K.  Top center:
$T_{in}=10^8$K.  Top right: $T_{in}=10^{7.5}$ K.  Bottom left:
$10M_\odot$ BH candidate with a surface, and a base temperature
$T_{in}=10^{7.5}$~K.  Bottom center: $T_{in}=10^7$ K.  Bottom right:
$T_{in}=10^{6.5}$ K.}  
\includegraphics[height=.7\textheight]{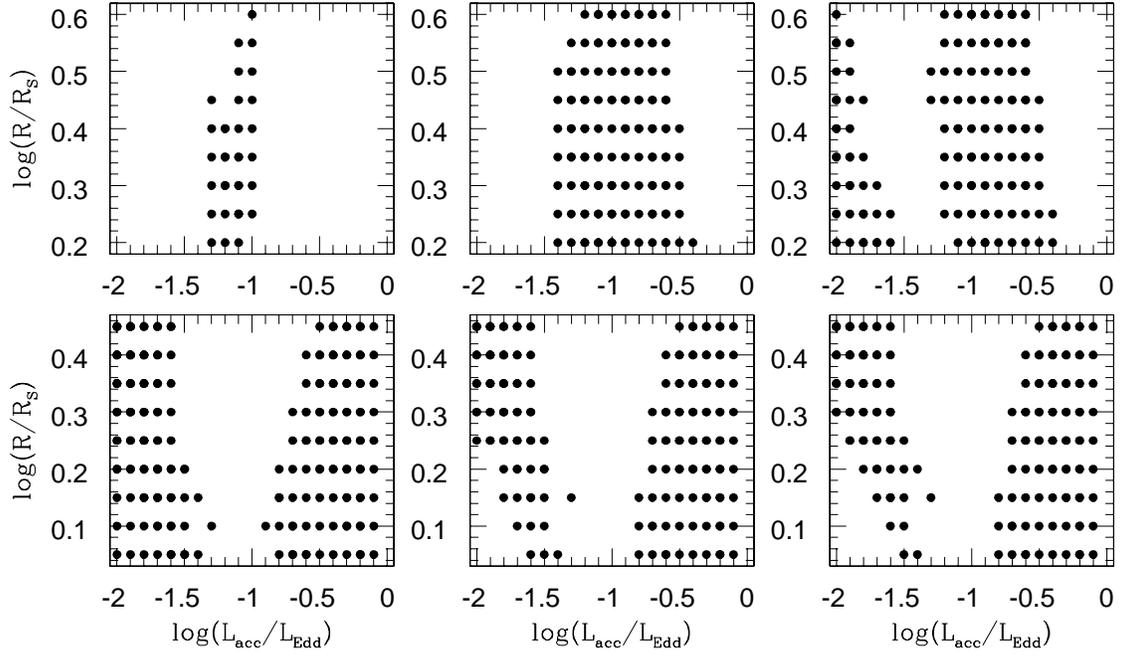}
\end{figure}

The top three panels of Fig. 1 show results for solar composition
material ($X_0=0.7$, $Y_0=0.27$, $Z_0=0.03$) accreting on a NS of mass
$1.4M_\odot$.  We consider a range of accretion rates, parameterized
by the ratio $L_{acc}/L_{Edd}$, where we take the Eddington luminosity
to be $L_{Edd}=4\pi GMc/\kappa_{es}$ with $\kappa_{es}=0.4 ~{\rm
cm^2\,g^{-1}}$.  We also consider a range of radii for the NS:
$\log(R/R_S)=0.2-0.6$ (corresponding to $R=6.5-16$ km).  For each
choice of $L_{acc}/L_{Edd}$ and $R/R_S$, we try three values of the
surface mass density of the accretion layer: $\Sigma_{max} = 10^9,
~10^{10}, ~10^{11} ~{\rm g\,cm^{-2}}$.  If any of the three cases is
unstable, i.e., if it has any eigenvalue with ${\rm Re}(\gamma)
>\gamma_{acc}$, we say that the system cannot achieve a stable steady
state and that it will exhibit Type I bursts.

The results presented in Fig. 1 correspond to three choices of the
temperature at the base of the accretion layer: $T_{in}=10^{8.5},
~10^8, ~10^{7.5}$~K (from left to right).  In the present work, we are
primarily interested in X-ray novae.  Since these sources have very
low luminosities in quiescence ($L_X<10^{33} ~{\rm erg\,s^{-1}}$ for
NS X-ray novae), the core temperatures of the NSs are likely to be
$T_{in} \ \sles\ 10^8$ K \cite{BBC02}.

The calculations shown in Fig. 1 indicate that NSs are unstable to
bursts for a wide range of $T_{in}$, but that the width of the
instability strip (in $L_{acc}$) is less for higher values of
$T_{in}$.  The reason for the latter is clear from the analysis in
\cite{P82} where it is shown that as the flux escaping from the
stellar core into the accretion layer increases (which happens when
$T_{in}$ increases), bursting behavior is restricted to a smaller
range of $\dot \Sigma$.

Assuming that $T_{in} \ \sles\ 10^8$ K, we see from Fig. 1 that NSs
should be unstable to bursts for accretion luminosities up to $L_{acc}
\approx 0.3 L_{Edd}$.  The prediction is generally consistent with
observations: the only NSs that are known not to burst are bright Z
sources with $L_{acc}\to L_{Edd}$ \cite{Metal95}, and X-ray pulsars.
Although X-ray pulsars are significantly less luminous than Eddington,
they accrete effectively at close to the Eddington rate since the
accreting matter is channeled onto a small area on the NS surface by
strong magnetic fields (see \cite{L00} for a detailed discussion of
this argument).

Immediately below $L_{acc}\sim 0.3L_{Edd}$, the instability is of a
mixed type in which a He-burning instability triggers a burst in which
He and H both burn explosively.  At lower luminosities, nearly all the
H is burned steadily, and the instability corresponds to a pure He
burst.  These results are consistent with previous work \cite{B98}.
Our calculations indicate that bursting behavior cuts off below an
accretion luminosity $L_{acc}\sim10^{-1.5}L_{Edd}$.  The cutoff is the
result of the restriction $\Sigma_{max}\le10^{11} ~{\rm g\,cm^{-2}}$
in the model.  Systems with luminosities below the cutoff are still
unstable to He bursts, but only at extremely high column density
$\Sigma_{max}$.  Correspondingly, the recurrence times $t_{rec}$ of
bursts are extremely long --- years in some cases according to our
calculations --- which makes these systems less useful for the present
study, since accretion outbursts of X-ray novae last no more than a
few months.

For $T_{in}=10^{7.5}$ K, there is a second instability strip at low
luminosities, around $L_{acc} \sim 10^{-2}L_{Edd}$.  This strip
corresponds to pure H bursts (the possibility of such bursts was first
noted in \cite{FHIR87}).  An interesting difference between the two
instability strips in this panel is that the strip on the right
generally has complex eigenvalues $\gamma$ for the unstable modes
while that on the left has real eigenvalues.

\subsection{Type I Bursts on Black Hole Candidates}

The bottom three panels in Fig. 1 show results for a $10M_\odot$ BH
candidate with a surface.  The three panels correspond to different
choices of the base temperature: $T_{in}=10^{7.5}, ~10^7, ~10^{6.5}$~K
(from left to right).  The particular choices of $T_{in}$ are
motivated by the extraordinarily low quiescent luminosities of many BH
X-ray novae ($L_X<10^{31} ~{\rm erg\,s^{-1}}$, see \cite{Getal01}; a
few BH novae are brighter than this, but even these are not likely to
have $T_{in} > 10^{7.5}$ K).  We consider stellar radii in the range
$\log(R/R_S)=0.05-0.45$, corresponding to $R$ between $(9/8)R_S$ and
$3R_S$.  The former limit is the smallest radius in general relativity
for an object whose density either decreases or remains level with
increasing radius \cite{ST83}.

The calculations indicate that BH candidates with surfaces are similar
to NSs in their bursting behavior.  Except for a modest rightward
shift of the positions of the instability zones, the results in the
lower three panels are quite similar to those for a NS with a
similarly low value of $T_{in}$ (upper right panel).  As in the case
of NSs, bursts are expected from BH candidates even for larger values
of $T_{in}$ than we have considered here, except that the instability
strip becomes narrower when $T_{in}$ exceeds $10^8$ K.  (As already
discussed, such high core temperatures are ruled out.)  We conclude
that BH candidates are as prone to Type I bursts as NSs are, provided
they have surfaces.

Because our model focuses only on the most important physical effects,
and neglects some details, the exact positions of the instability
strips in Fig. 1 may be uncertain at the level of say a factor of two
in accretion luminosity.  We believe, however, that the overall
pattern of instability is fairly robust.  If at all, our computations
are likely to be conservative in the sense that bursting behavior may
be even more widespread than we predict.

\section{Discussion}

It is clear from the theory of bursts (\cite{B98} and references
therein) that bursting behavior is largely determined by the surface
gravity $g$, the mass accretion rate, and the composition of the
accreting material.  It is also clear that the accreting object must
have a surface in order to burst.  If there is no surface, material
cannot accumulate, and therefore cannot become hot or dense enough to
trigger a thermonuclear explosion.  Indeed, using this argument,
astronomers routinely identify any X-ray binary that has a Type I
burst as a NS system.

We suggest that the argument could equally well be applied in reverse,
at least in the case of X-ray novae: If an X-ray nova with a BH
candidate (identified independently via a mass measurement) does not
exhibit Type I bursts at any time during its accretion outburst, then
we argue that the primary star cannot have a surface --- the object
must be a BH with an event horizon.  To our knowledge, this simple
argument has not been explored so far in the literature \citep[but
see][]{M01}.

While the argument is simple, it is not obvious that it is correct.
The reason is that not all systems with surfaces necessarily have
bursts; for instance, many NS X-ray binaries, e.g., the brighter Z
sources and X-ray pulsars, do not burst.  Clearly, it is necessary to
understand through physical modeling under what conditions a compact
star with a surface will or will not burst.  This is the motivation
behind the calculations presented here.  Our results are shown in
Fig. 1, with the top three panels corresponding to NSs and the bottom
three panels to $10M_\odot$ BH candidates with surfaces. In brief, we
find that BH candidates with surfaces ought to burst as prolifically
as NSs.

A notable feature of our results is that we find bursting behavior in
BH candidates (with surfaces) for quite a wide range of
$L_{acc}/L_{Edd}$, regardless of the radius of the accreting star.  As
discussed in the Introduction, the accretion rate in a typical X-ray
nova spans a wide range during an accretion outburst.  At maximum, the
sources often approach $L_{acc}\sim L_{Edd}$, at the right edge of a
given panel in Fig. 1, while in quiescence, the sources are extremely
underluminous, well off the left edge of the panels.  Thus, during the
rise to outburst and subsequent decline, each BH X-ray nova enters the
plot on the left, moves horizontally all the way across, and then
moves horizontally back, to disappear off the left edge.  The journey
typically takes a few months.  During this entire journey, it is an
amazing observational fact that not one of the BH candidates has ever
had a Type I burst.  Our calculations indicate that this is impossible
if the sources have surfaces.  Regardless of the radius of the BH
candidate's surface (which is of course unknown), the trajectory that
the star traverses during an accretion outburst would always intersect
wide stretches of instability.  Moreover, according to our estimates,
the burst recurrence time is of the order of a day in many regions of
the diagram, which means that each source should exhibit many bursts
during its months-long journey.  The absence of bursts in BH X-ray
novae is thus highly significant and argues strongly for the lack of
surfaces in these systems.

Before we can claim that this ``proves'' the reality of the event
horizon, more work is needed.  

First, we need to explain why some NSs do not have bursts, and we
should provide a convincing argument why the same explanation cannot
be applied to BH candidates.  We have discussed in this paper two
categories of NSs that do not burst.  (i) Most Z sources do not burst,
and we explained this as the result of their luminosities being too
close to $L_{Edd}$.  Since BH X-ray novae span a wide range of
$L_{acc}$, the explanation obviously does not apply to them. (ii)
X-ray pulsars generally do not burst, and in this case the standard
explanation is that magnetic funneling causes the ``effective''
accretion rate to approach the Eddington level locally.  Since BH
X-ray novae do not show coherent pulsations, it is extremely unlikely
that they have the kind of magnetic funneling present in X-ray
pulsars; therefore, once again, the explanation cannot be invoked for
BH X-ray novae.  We are not aware of other classes of NSs that do not
burst, but if any are identified, it is necessary to explain the lack
of bursts through physical modeling.  It is also necessary to
demonstrate that the same explanation does not apply to BH X-ray
novae.

Second, we need to show that the model is able to reproduce the main
features of Type I bursts as observed in NS X-ray binaries.  For
instance, the statistics of burst durations and recurrence times
\cite{vPPL88} 
ought to appear naturally in a realistic model.

Third, the role of the inner boundary condition needs to be explored
in detail.  In Fig. 1, we see that different choices of the base
temperature for a BH candidate give similar results.  We have tried
other boundary conditions (e.g., we have applied a boundary condition
on the flux \cite{P82} instead of the temperature), and also tried
changing the composition of the accreting gas.  In all cases we find
that the accumulating layer is unstable to bursts for a wide range of
$L_{acc}/L_{Edd}$, both for $1.4M_\odot$ NSs and $10M_\odot$ BH
candidates with surfaces.

Fourth, the difficult issue of flame propagation over the surface of
the star once the instability has been triggered needs to be addressed
(e.g., \cite{SLU02}).  In this context, we note the empirical fact
that the flame clearly propagates with no trouble on NSs, as indicated
by the fact that these objects do exhibit bursts.  The key question
is: could the propagation physics be so delicate that the burning
front stops propagating when the mass of the star changes from
$1.4M_\odot$ to $10M_\odot$?  We do not know the answer to this
question.

We should caution, morever, that the analysis presented here assumes
that the accumulating gas on the surface of a BH candidate behaves
like normal matter, with nucleons and electrons.  Obviously, our
arguments would become invalid if the properties of the gas changed
drastically, e.g., if the nuclei disappeared and were replaced by
quarks.  Whether such extreme changes are plausible on the surface of
a strange star remains to be seen.  The density and pressure at the
base of the bursting layer do not go above ${\rm few}\times10^8 {\rm
g\,cm^{-3}}$ and $10^{26} ~{\rm erg\,cm^{-3}}$, respectively, even in
the most extreme cases we have considered.  It is hard to imagine
exotic physics being important under these conditions \cite{G97}.

On the observational front, we should check whether some NSs that lie
within the unstable regions of Fig. 1 are stable to bursts.  Any
obvious large-scale disagreement between the observed burst behavior
of NSs and the results presented here would indicate that the model is
missing important physics.  We are not aware of serious discrepancies
of this nature, but the matter clearly deserves careful study.

In the case of BH X-ray novae, we should use observational data to
derive strict quantitative limits on bursting activity.  The current
data are already compelling --- no burst has been reported from any of
the dozen or so BH X-ray novae with dynamically measured primary
masses.  Nevertheless, a careful study of archival data on past X-ray
novae as well as data to be collected on future novae is needed to
rule out the possibility that bursts might have been missed.  This is
well worth the effort since a firm demonstration that BH transients do
not have Type I bursts at any point in their light curves would, as we
have shown, provide a strong argument for the presence of event
horizons in these systems.


\begin{theacknowledgments}
We thank Andrew Cumming, Alex Ene, Kristen Menou, Bohdan Paczynski and
Greg Ushomirsky for useful discussions.  RN was supported in part by
the W. M. Keck Foundation as a Keck Distinguished Visiting Professor.
RN's research was supported by NSF grant AST-9820686 and NASA grant
NAG5-10780.  JSH was supported by the Chandra Postdoctoral Fellowship
Award \# PF0-10015 issued by the Chandra X-ray Observatory Center,
which is operated by the Smithsonian Astrophysical Observatory for and
on behalf of NASA under contract NAS8-39073.
\end{theacknowledgments}


\bibliographystyle{aipproc}   

\bibliography{paper}

\IfFileExists{\jobname.bbl}{}
 {\typeout{}
  \typeout{******************************************}
  \typeout{** Please run "bibtex \jobname" to optain}
  \typeout{** the bibliography and then re-run LaTeX}
  \typeout{** twice to fix the references!}
  \typeout{******************************************}
  \typeout{}
 }

\end{document}